\documentclass{PoS}
\usepackage{dsfont}

\title{Numerical simulation of self-dual U(1) lattice field theory with electric and magnetic matter\thanks{Based on a parallel talk by 
M.~Anosova.}}

\ShortTitle{Self-dual U(1) lattice field theory with electric and magnetic matter}

\author{Maria Anosova$^1$, Christof Gattringer$^2$\footnote{On leave of absence from University of Graz, 8010 Graz, Austria.} , 
Nabil Iqbal$^3$\footnote{NI is supported in part by the STFC
under consolidated grant ST/L000407/1.} , 
Tin Sulejmanpasic$^3$ \footnote{TS is supported by the Royal Society University Research Fellowship.}\\

\vskip2mm

$^1\;$Institute of Physics\footnote{Member of NAWI Graz.} , University of Graz, 8010 Graz, Austria

$^2\;$FWF - Austrian Science Fund, 1090 Vienna, Austria 

$^3\;$Department of Mathematical Sciences, Durham University, Durham DH1 3LE, UK

\vskip2mm
        
E-mail: \email{mariia.anosova@uni-graz.at} \\ 
E-mail: \email{christof.gattringer@fwf.ac.at} \\
E-mail: \email{nabil.iqbal@durham.ac.uk}  \\
E-mail: \email{tin.sulejmanpasic@gmail.com}  
}

\abstract{We study a recently proposed formulation of U(1) lattice field theory with electric and magnetic matter based on the Villain
formulation. This discretization allows for a duality that gives rise to relations between weak and strong gauge coupling. There 
exists a self-dual value of the gauge coupling where one may study the model as a function of the remaining matter coupling. 
Using Monte Carlo simulations based on a worldline/worldsheet representation of the system we evaluate order parameters 
for spontaneous breaking of self-duality. We find that in some interval of the matter coupling self-duality becomes broken 
spontaneously. We determine the endpoints of this interval and study the nature of the corresponding critical points. Finally 
we explore the system away from the self-dual gauge coupling and show that when crossing the self-dual point 
a first order jump is seen in the order parameters.}

\FullConference{38th International Symposium on Lattice Field Theory - Lattice2021\\
		July 26 -  30, 2021\\
		Zoom/Gather@MIT}

\begin{document}

\section{Introduction}

\noindent
U(1) gauge theories are essential in our everyday experience and, along with gravity, influence our world in a more direct way than the other two fundamental forces. Moreover, emergent U(1) gauge theories are omnipresent in condensed matter systems. Yet U(1) gauge theories in four spacetime dimensions are thought not to be genuine quantum field theories. They are not asymptotically free, and hence their definition requires embedding them in another, UV complete theory, such as a non-abelian gauge theory or a lattice gauge theory. 

In four spacetime dimensions in a continuum description, free U(1) gauge theories have two global symmetries, because $\partial^\mu J_{\mu\nu}= \partial ^\mu\tilde J_{\mu\nu}=0$, where $J_{\mu\nu}\propto F_{\mu\nu}, \, \tilde J_{\mu\nu}\propto \epsilon_{\mu\nu\rho\sigma}F^{\rho\sigma}$, which we will refer to as \emph{electric} and \emph{magnetic} symmetries\footnote{For free U(1) gauge theory, both of these symmetries are U(1) symmetries. When matter fields are added, these symmetries may be reduced.}. Such symmetries, resulting in two-antisymmetric-index currents are referred to as 1-form symmetries\footnote{The electric symmetry $J_{\mu\nu}=F_{\mu\nu}$ is always a 1-form symmetry in any dimension, but the magnetic symmetry is a $d-3$-form symmetry, where $d$ is the number of space-time dimensions.} \cite{Gaiotto:2014kfa}. These conservation laws can be thought of as the Gauss law for electric and magnetic field: the electric and magnetic flux are independent of the shape of the Gauss surface, and only depend on the charge within\footnote{Note the distinction between these conservation laws and the usual N\"other charges. The ordinary, or 0-form symmetries, have charges which are integrals over space, i.e., they are codimension 1 operators. In contrast, a $p$-form conserved charge is a codimension $p+1$ operator.}. 

It is a common lore that UV completions of a U(1) gauge theory always requires monopoles \cite{Polchinski:2003bq}. Indeed both embeddings, the one via non-abelian gauge theory, as well as Wilson lattice gauge theory result in a system which has dynamical monopoles. But that means that there is no magnetic symmetry in the underlying system. However, some condensed matter systems are believed to be effectively described by a U(1) gauge theory where only monopoles of non-minimal charge are dynamical, and some discrete part of the magnetic symmetry survives. This leads to the question whether gauge theories can be formulated where monopole matter is absent, or where monopoles are treated on equal footing with electric matter (e.g., they can be endowed with spin, bare mass, or short range interactions). Indeed, it is well known that the continuum formulation of U(1) gauge theory enjoys an electric-magnetic duality, and at least in principle electric and magnetic matter can be treated on equal footing (see, e.g., \cite{QFT_for_math}). In fact U(1) lattice gauge theories were often used where monopole particles were identified in lattice configurations, and suppressed or given an action to \cite{Barber:1984ak,Kerler:1994qc,Kerler:1996cr,Motrunich:2003fz}, but such procedures treat magnetic and electric matter content in a very different way, and at the very least obscure the formal equivalence between electric and magnetic matter.

On the other hand, Villain's lattice action \cite{Villain:1974ir} is known to have better self-dual features, and the 
EM self-duality was initially discovered, not in the continuum but in specific lattice systems \cite{Peskin:1977kp,Cardy:1981qy,Cardy:1981fd,Shapere:1988zv}. Yet it was noted only recently in \cite{Sulejmanpasic:2019ytl,WIP1,WIP2}, that a certain modification of Villain's action\footnote{A similar construction was recently used in lattice models of fractons \cite{Gorantla:2021svj}.} can be used to construct models where magnetic and electric matter can be coupled on equal footing, and do not appear as artifacts of a theory. The gist of it is that such a formulation features both electric and magnetic gauge fields in the lattice action, which allows one to endow both electric and magnetic matter with spin, internal quantum numbers and arbitrary gauge charge, or to eliminate them all together, by setting the corresponding matter action to zero. This raises the question of whether such U(1) lattice gauge theories can be considered as genuine quantum field theories. Indeed, based on symmetries alone, such theories can have more symmetries than non-abelian gauge theories, whose global symmetry group is typically limited to the electric $\mathbb Z_N$ center symmetry and electric flavor symmetry\footnote{One may gauge a subgroup of the discrete center symmetry and thereby obtain a bonus magnetic center symmetry, see e.g. \cite{Gaiotto:2017yup}.}. In contrast, the modified Villain gauge theories can be endowed with $Z_Q$ magnetic center symmetries (by setting all magnetic charges to multiples of the integer $Q$) and/or magnetic flavor symmetries, by coupling multiple flavors. This raises the question whether any of these theories 
have a continuum limit. 
Moreover, if the actions of the electric and magnetic matter are identical, and if the gauge coupling is dialed to a particular value, the lattice model becomes \emph{exactly} self-dual\footnote{More precisely the self-duality generator actually does not exponentiate to unity, but to charge-conjugation + lattice translation by one unit in all directions. In a continuum limit, one can can think of the self-dual symmetry generator $S$ as forming a $\mathbb Z_4$ group, where $S^2=C$ is the charge conjugation operator.}.

In this contribution we explore duality and self-duality in the simplest of such theories: a U(1) gauge theory with a single scalar electric matter field and a single scalar magnetic matter field. For a generic electric coupling $e={1}/{\sqrt{\beta}}$, the system has no global symmetry. However, duality maps the lattice action to itself, with dual coupling $\beta\rightarrow \widetilde{\beta} = 1/ 4\pi^2 \beta$. If 
$\beta$ is chosen to be $\beta=\beta^* \equiv 1/2\pi$, the self-duality transformation is a symmetry of the theory\footnote{Note that, unlike Kramers-Wannier duality of the 2d Ising system, the self-duality of the model we discuss is a genuine symmetry at the self-dual point. The point is that when an Ising system is dualized, the resulting dual theory is not exactly the original theory, but an original theory coupled to a topological quantum field theory -- roughly where its global symmetry group is gauged. For more on why Kramers-Wannier duality is not a genuine duality, see \cite{Kapustin:2014gua}.}. At this point one may study the system as a function of the remaining 
matter coupling $J$ (which for full self-duality has to be the same for electric and magnetic matter). 

\begin{figure}[t!] %  figure placement: here, top, bottom, or page
   \centering
   \includegraphics[width=4in,clip]{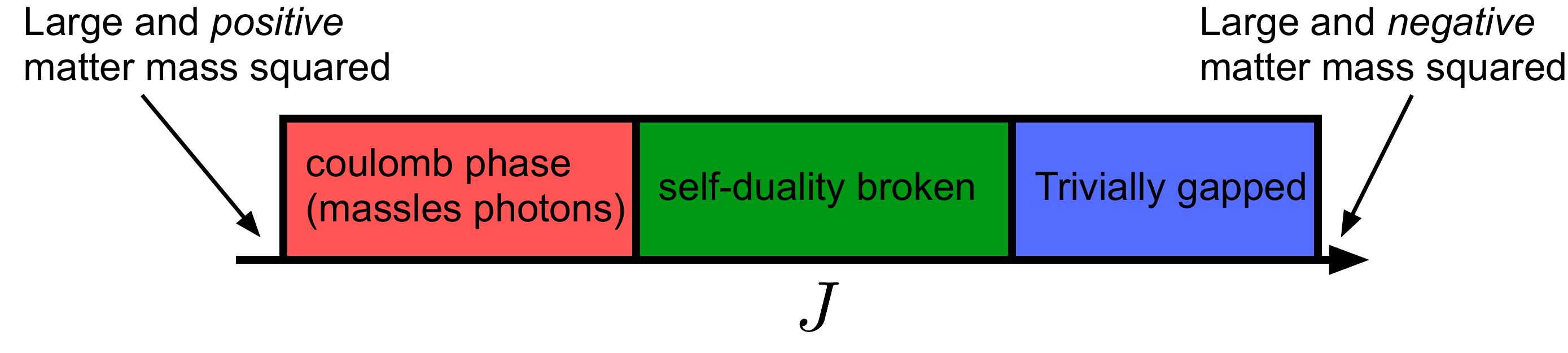} 
   \vspace{-6mm}
   \caption{The phase diagram of U(1) lattice gauge theory with electric and magnetic matter.}
   \label{fig:phase_diag_selfdual}
\end{figure}

We here present our first numerical results from a Monte Carlo simulation of this system at the self-dual coupling $\beta^*$ with 
varying matter coupling $J$ (see below for the definition and the explicit formulation of the model). For $J=0$  matter decouples completely (think of it as the limit $m^2\rightarrow \infty$, where $m$ is the bare mass) and we just have a Coulomb (i.e., photon) phase. As $J$ is increased, electric and magnetic matter want to condense, but there is a tension, because the condensation of electric (magnetic) matter, confines the magnetic (electric) matter preventing it from condensing. So we expect that, for some $J=J_1$ we enter a phase of coexistence between electric and magnetic condensation, and hence a spontaneous breaking of the self-dual symmetry ensues. If $J$ is increased even further, we enter a deeply Higgsed lattice regime, which can be solved exactly for our model, and is trivially gapped, i.e., electric condensation (i.e., a Higgs phase) and magnetic condensation (i.e., a confinement phase) exhibit the Fradkin-Shenker continuity \cite{Fradkin:1978dv}. So we expect that at some $J=J_2$ the self-dual symmetry gets restored again. We summarize the phase diagram in Fig.~\ref{fig:phase_diag_selfdual}. 

Using suitable order parameters for self-duality we find that self-duality is indeed broken in some range of $J$ and we determine
and analyze the corresponding (critical) endpoints. Varying the gauge field coupling around the self-dual value $\beta^*$ 
we find that the order parameters exhibit a first order jump at $\beta^*$. 
Our results constitute one of the first non-perturbative instances of spontaneous breaking of self-duality\footnote{See \cite{Somoza:2020jkq} for a recent study of self-duality breaking in a 3d model of a discrete gauge theory.}, thus adding another 
aspect to the rich phenomenology of duality relations in quantum field theory. 

\section{Definition of self-dual U(1) lattice gauge theory with electric and magnetic matter}

\noindent
The gauge field action is based on the Villain formulation \cite{Villain:1974ir}  and is given by
\begin{equation}
S_g[A^e\!,\,n]  \; \equiv \; \frac{1}{2} \sum_{x} \! \sum_{\mu < \nu} ( F_{x,\mu \nu} )^2 \; ,
\label{gauge_action}
\end{equation}
where the sum runs over all plaquettes of our lattice. The field strength tensor is defined as 
$ F_{x,\mu \nu}  \equiv (d A^e)_{x,\mu \nu} + 2 \pi \, n_{x,\mu \nu}  = (d A^e \; + \; 2 \pi \, n)_{x,\mu \nu}$,
with the exterior derivative defined as 
$(d A^e)_{x,\mu \nu}  \equiv  A^e_{x + \hat{\mu},\nu} - A^e_{x,\nu} - A^e_{x + \hat{\nu},\mu} + A^e_{x,\mu}$. 
The \emph{electric gauge fields} $A^e_\mu(x) \in [-\pi, \pi]$ are assigned to the links of our 4-d lattice, while the
\emph{Villain variables} $n_{x,\mu \nu} \in \mathds{Z}$
live on the plaquettes. The lattice we consider has size $N_s^3 \times N_t$ and we use periodic
boundary conditions for all directions. 

We will couple electric matter using the link variables defined as $U_\mu(x) = e^{\, i A^e_\mu(x)}$. Clearly the link variables are 
invariant under shifts of $A^e_\mu(x)$ by multiples of $2 \pi k_\mu(x)$ with $k_\mu(x) \in \mathds{Z}$, 
while the exterior derivatives $(d A^e)_{x,\mu \nu}$ are not. The Villain variables 
$n_{x,\mu \nu}$ may be considered as gauge fields for the shift symmetry and summing them over all integers renders 
the field strength $F_{x,\mu \nu}$ invariant under the shift symmetry. 

Note, however, that we may impose an additional constraint on the Villain variables \cite{Sulejmanpasic:2019ytl,WIP1}, since the 
contribution from the shift alters the exterior derivative $(d A^e)_{x,\mu \nu}$ by the term $(d k)_{x,\mu \nu}$, and this term
obeys $d (d k)_{x,\mu \nu} = 0$, due to the nilpotency $d^2 = 0$ of the exterior derivative (for a brief summary
of the necessary tools and conventions for lattice differential forms see the appendices of  
\cite{Sulejmanpasic:2019ytl,WIP1}, as well as \cite{Seiler:1982pw,Wallace}). Thus we may 
impose the \emph{closedness constraint}
\begin{equation}
(dn)_{x,\mu \nu \rho}  =  0 \;\;\; \forall  \; (x, \mu \nu \rho) \; , \; 
(dn)_{x,\mu \nu \rho}  \equiv 
n_{x + \hat \mu, \nu \rho} - n_{x, \nu \rho} - 
n_{x + \hat \nu, \mu \rho} + n_{x, \mu \rho} +
n_{x + \hat \rho, \mu \nu} - n_{x, \mu \nu} ,
\label{closedness}
\end{equation}
which implements vanishing exterior derivative $(dn)_{x,\mu \nu \rho}$ of the Villain variables on all cubes 
$(x,\mu \nu \rho) $ of the lattice. The physical role of the closedness constraint is to remove magnetic monopoles
from our lattice formulation. The closedness constraint can be formulated as a product of Kronecker deltas (our notation here is
$\delta(n) = \delta_{n,0}$),
\begin{equation}
\prod_x \prod_{\mu < \nu < \rho} \delta \big( (dn)_{x,\mu \nu \rho} \big) \; = \; 
\prod_x \prod_{\mu < \nu < \rho} \int_{-\pi}^\pi \frac{d A^m_{x,\mu\nu\rho}}{2\pi} \, 
e^{\, i A^m_{x,\mu\nu\rho} (dn)_{x,\mu \nu \rho}} \; ,
\end{equation}
where in the second step we have written the Kronecker deltas with an integral representation which introduces the 
\emph{magnetic gauge fields}  $A^m_{x,\mu\nu\rho} \in [-\pi,\pi]$ assigned to the cubes of the lattice. 

We may now combine the gauge field action and the constraints into the joint Boltzmann factor
\begin{equation}
B_\beta[A^e,A^m]  \; \equiv \; \sum_{\{ n \}} \; e^{ \, - \beta \, S_g[A^e\!,\,n] } 
\; e^{\, i \, \sum_x \! \sum_{\mu < \nu < \rho} A^m_{x,\mu \nu \rho} (dn)_{x,\mu \nu \rho}}  \; ,
\label{boltzmann_both}
\end{equation}
where $\beta$ is the (inverse) gauge coupling $\beta = 1/e^2$, with $e$ the electric charge. By $\sum_{\{ n \}}$ we denote the
sum over all configurations of the Villain variables,
\begin{equation}
\sum_{\{ n \}}  \; \equiv \; \prod_{x} \prod_{\mu < \nu} \sum_{n_{x,\mu \nu} \in \mathds{Z}} \; , \; \; 
\int \! D[A^e] \; \equiv \; \prod_{x} \prod_{\mu} \int_{-\pi}^\pi \! \! \frac{d A^e_{x,\mu}}{2 \pi} \; , \; \; 
\int \! D[A^m] \; \equiv \; \prod_{x} \prod_{\mu< \nu < \rho} \int_{-\pi}^\pi \! \! \frac{d A^m_{x,\mu,\nu\rho}}{2 \pi} \; ,  
\label{measures}
\end{equation}
where we have also introduced the measures over the electric and the magnetic gauge fields. Using these we may write 
the partition function as
$Z(\beta) = \int \!\! D[A^e] \int \!\! D[A^m] \; B_\beta[A^e,A^m]$,
which constitutes a formulation of 4-d U(1) lattice gauge theory that is free of monopoles. 

Using Poisson resummation and switching to the dual lattice, one may show \cite{Sulejmanpasic:2019ytl,WIP1} that our 
lattice discretization of U(1) gauge fields is also self-dual, i.e., the partition sum obeys ($V = N_s^3 N_t$)
\begin{equation}
Z(\beta) \; = \, \left(\!\frac{1}{2\pi \beta}\!\right)^{\!\!3V} \!\! Z(\widetilde{\beta}) \qquad \mbox{with} \qquad 
\widetilde{\beta} \; = \; \frac{1}{4 \pi^2 \beta} \; .
\label{duality_Z}
\end{equation} 
The self-duality relation connects the weak and strong coupling regimes of the theory.  

From a physical point of view the key step to self-duality is the removal of lattice monopoles by augmenting the Villain
action with the closedness constraint (\ref{closedness}). There is an interesting option to generalize the pure lattice 
gauge theory considered so far in a self-dual way by abandoning the closedness constraint and explicitly coupling 
magnetic matter and its dual counterpart,
i.e., electric matter. For the electric matter fields we here use U(1)-valued spins 
$\phi^e_x$ attached to the sites which we parameterize as $\phi^e_x = e^{\, i \varphi_x^e}$ with $\varphi_x^e \in [-\pi, \pi]$. We 
remark that it is straightforward to replace the U(1)-valued matter by a general charged bosonic or fermionic field. 
The partition sum for the electric matter in a background configuration of the electric gauge field $A^e_\mu(x)$ is given by
\begin{equation}
Z[A^e,J_e] \; \equiv \;  \int \!\! D[\phi^e] \, e^{ \, J_e S_e[\phi^e,\,A^e]} \; , \; \; 
\int \!\! D[\phi^e] \; \equiv \; \prod_x \int_{-\pi}^\pi \frac{d \varphi_x^e}{2 \pi} \; ,
\label{Zelectric}
\end{equation}
with the action for electric matter defined as
\begin{equation}
S_e[\phi^e,A^e] \; \equiv \; \frac{1}{2}  \sum_{x,\mu} 
\left[ \phi_x^{e\, *} \,  e^{i A_{x,\mu}^e} \, {\phi_{x+\hat{\mu}}^e} + c.c. \right] 
\; = \; 
\sum_{x,\mu} \cos \big( \varphi_{x+\hat{\mu}}^e - \varphi_x^e + A_{x,\mu}^e \big) \; .
\end{equation}
The partition sum contains a new parameter, the \emph{electric matter coupling} $J_e \in \mathds{R}_+$.

In a similar way we now define the partition sum for magnetically charged matter in the background of the magnetic gauge field. 
The magnetic gauge field $A^m_{x,\mu \nu \rho}$ is attached to the 3-cubes of the lattice. It is natural to switch to the dual lattice and
by $\widetilde{A}_{\tilde{x},\mu}^m$ we denote the magnetic gauge fields now labelled with the coordinates $(\tilde{x},\mu)$ for the 
links of the dual lattice (see again the appendices of \cite{Sulejmanpasic:2019ytl,WIP1} for our conventions). 
The magnetically charged scalar field $\widetilde{\phi}_{\tilde{x}}^m \in$ U(1) is assigned to the sites of the dual lattice and we 
parameterize it in the form $\widetilde{\phi}_{\tilde{x}}^{m} = e^{\, i \widetilde{\varphi}_{\tilde{x}}^{\, m}}$, 
where $\widetilde{\varphi}_{\tilde{x}}^{\, m} \in [-\pi, \pi]$.
The corresponding partition function is identical to the partition function (\ref{Zelectric}) for the electric matter
but now is defined entirely on the dual lattice,
\begin{eqnarray}
\widetilde{Z}\big[\widetilde{A}^m,J_m\big] & \equiv &  \int \!\! D\big[\widetilde{\phi}^m\big] \, 
e^{\, J_m \widetilde{S}_m \big[\widetilde{\phi}^m,\, \widetilde{A}^m \big]} \; , \; \; 
\int \!\! D \big[\widetilde{\phi}^m\big] \; \equiv \; \prod_{\tilde{x}} \int_{-\pi}^\pi 
\frac{d \widetilde{\varphi}_{\tilde{x}}^{\,m}}{2 \pi} \; ,
\label{Zmagnetic}
\\
\widetilde{S}_m\big[\widetilde{\phi}^m,\widetilde{A}^m\big] & \equiv &  \frac{1}{2}  
\sum_{\tilde{x},\mu} \left[  
\widetilde{\phi}_{\tilde{x}}^{m \, *} \,  e^{i \widetilde{A}_{\tilde{x},\mu}^m}  \, \widetilde{\phi}_{\tilde{x}+\hat{\mu}}^m 
+ c.c. \right] \; = \;   
\sum_{\tilde{x},\mu} \cos \big( \widetilde{\varphi}_{\tilde{x}+\hat{\mu}}^{\,m}
- \widetilde{\varphi}_{\tilde{x}}^{\,m}  + \widetilde{A}_{\tilde{x},\mu}^{\,m} \big) 
\; ,
\end{eqnarray}
and again we introduced a coupling parameter, here denoted as $J_m$. The overall partition sum is then obtained as 
$Z(\beta, J_e,J_m) \equiv \int \!\! D[A^e] \int \!\! D[A^m] \; B_\beta[A^e,A^m] \;
Z [A^e,J_e] \; \widetilde{Z}\big[\widetilde{A}^m,J_m\big]$,
and is a function of the gauge coupling $\beta$, as well as the electric and the magnetic matter couplings $J_e$ and $J_m$.

A duality relation also holds for the full theory with electric and magnetic matter \cite{Sulejmanpasic:2019ytl,WIP1},
\begin{equation}
Z(\beta, J_e, J_m) \; = \; \left(\!\frac{1}{2\pi \beta}\!\right)^{\!\!3V}  \,
Z\big(\widetilde{\beta}, \widetilde{J}_e, \widetilde{J}_m\big) 
\quad \mbox{with} \quad 
\widetilde{\beta} \; = \; \frac{1}{4\pi^2 \beta} \; , \quad 
\widetilde{J}_e \; = \; J_m  \; , \quad 
\widetilde{J}_m  \; = \; J_e  \; .
\label{Zfullduality}
\end{equation}
The key goal of this study is to explore this duality relation using numerical Monte Carlo simulations. Suitable observables for 
such a study may be obtained by derivatives of $\ln Z$ with respect to the couplings. In particular we consider the action densities, 
\begin{eqnarray}
\hspace*{-4mm} 
\langle F^2 \rangle_{\beta, \, J_e, \, J_m} & \equiv & - \, \frac{1}{3V} \frac{\partial}{\partial \beta} \ln \, Z (\beta, J_e, J_m)
\; \; \mbox{with} \quad F^2 \equiv \!\!  \sum_{x,\mu < \nu} \!\! (F_{x,\mu \nu})^{2}/6V = S_g[A^e\!,\,n] / 3V \; ,
\label{obs_F}
\\
\hspace*{-4mm} 
\left\langle s_e \right\rangle_{\beta, \, J_e, \, J_m} & \equiv & 
\frac{1}{4V} \frac{\partial }{\partial J_e} \, \ln \, Z (\beta, J_e, J_m)
\quad \, \mbox{with} \quad  s_e \equiv S_e/4V \; ,
\label{obs_se}
\\
\hspace*{-4mm} 
\left\langle \widetilde{s}_m \right\rangle_{\beta, \, J_e, \, J_m} & \equiv & 
\frac{1}{4V} \frac{\partial }{\partial J_m} \, \ln \, Z (\beta, J_e, J_m)
\quad \mbox{with} \quad  \widetilde{s}_m \equiv \widetilde{S}_m/4V \; .
\label{obs_sm}
\end{eqnarray}

Before we come to the discussion of our results we need to briefly address the actual simulation. Obviously the Boltzmann factor 
$B_\beta[A^e,A^m]$ in Eq.~(\ref{boltzmann_both}) is complex and the corresponding complex action problem prevents 
a direct application of Monte Carlo techniques. In order to overcome this problem one may
switch to a dual representation in terms of worldlines and worldsheets where the weight factors are real and positive. We do not
discuss the derivation of this representation or the corresponding Monte Carlo techniques here, but refer to 
\cite{Sulejmanpasic:2019ytl,WIP1,WIP2,Anosova:2019quw} for details. 

\section{Numerical results for simulations of the self-dual theory}

\noindent
In this contribution we report about our first results for simulations at the self dual-point, i.e., we set 
$\beta = \widetilde{\beta} = \beta^* = 1/2\pi$ and $J_e = J_m = J$, where $J$ is the only remaining variable coupling. 
We evaluate our observables following (\ref{obs_F}) --  (\ref{obs_sm}) and use
$\partial / \partial \beta = (\partial \widetilde{\beta} / \partial \beta) \, \partial/\partial \widetilde{\beta} = 
- 1/ 4\pi^2 \beta^2 \,  \partial/\partial \widetilde{\beta}$ for the derivative of the rhs.\ of the duality relation (\ref{Zfullduality})
with respect to $\beta$. After a few lines of algebra this leads to the self-duality relation for the expectation 
value of the gauge field action density,
\begin{equation}
\langle F^2 \rangle_{\beta^*\!, \, J}  \;  = \; \pi \; \; \; \; \forall \; J \; .
\label{F2constant}
\end{equation}
In other words, self-duality implies that $\langle F^2 \rangle_{\beta^*\!, \, J}$ is constant for all $J$.  
In a similar way one may apply derivatives with respect to the matter couplings $J_e,J_m$ on the two sides of 
(\ref{Zfullduality}) and then set $J_e = J_m = J$ afterwards to come up with a self-duality relation for the matter field action densities,
\begin{equation}
\left\langle s_e \right\rangle_{\beta^*\!, \, J} \; = \; 
 \left\langle \widetilde{s}_m \right\rangle_{\beta^*\!, \, J} \; \; \; \; \forall \; J \; .
\label{duality_phi2_sd}
\end{equation}
In order to explore spontaneous breaking of self-duality we thus may define the order parameters 
\begin{equation}
M_g \; \equiv \;  | F^2 \; - \; \pi | \qquad \mbox{and} \qquad M_m \; \equiv \; | s_e \, - \, s_g | \; .
\label{orderparam}
\end{equation}
If the vacuum expectation values of these order parameters are non-vanishing this signals that 
self-duality is broken spontaneously. 
We remark that absolute values were introduced in (\ref{orderparam}), such that we can study the effects of spontaneous 
symmetry breaking also on a finite lattice. When studying the order parameters with explicit 
breaking, i.e., for $\beta \neq \beta^*$, we may omit the absolute values. In addition we also consider the corresponding Binder cumulants to analyze potential critical behavior. 

\begin{figure}[t!!] %  figure placement: here, top, bottom, or page
   \hspace*{0mm}
   \includegraphics[width=80mm]{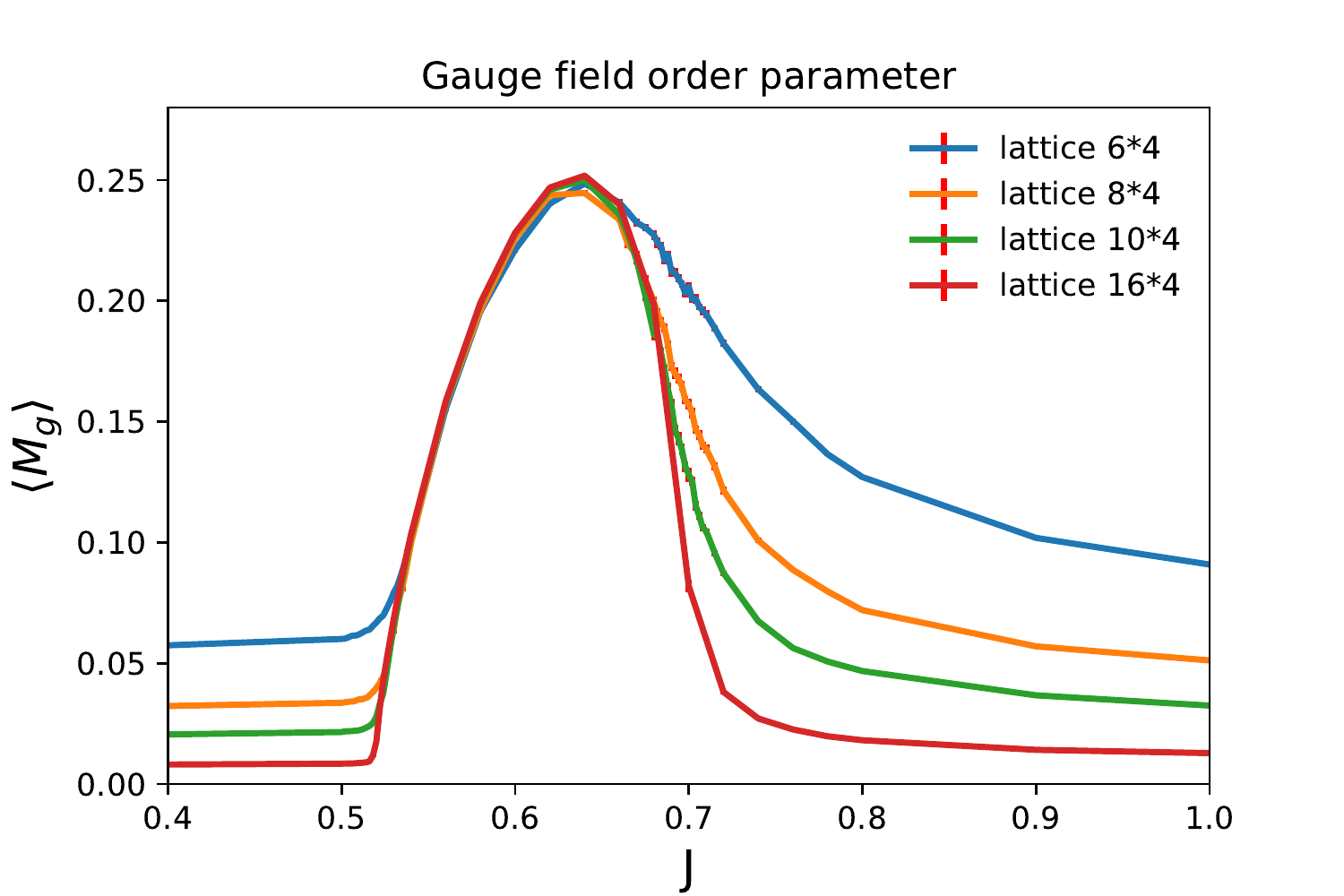} 
   \hspace*{-5mm}
   \includegraphics[width=80mm]{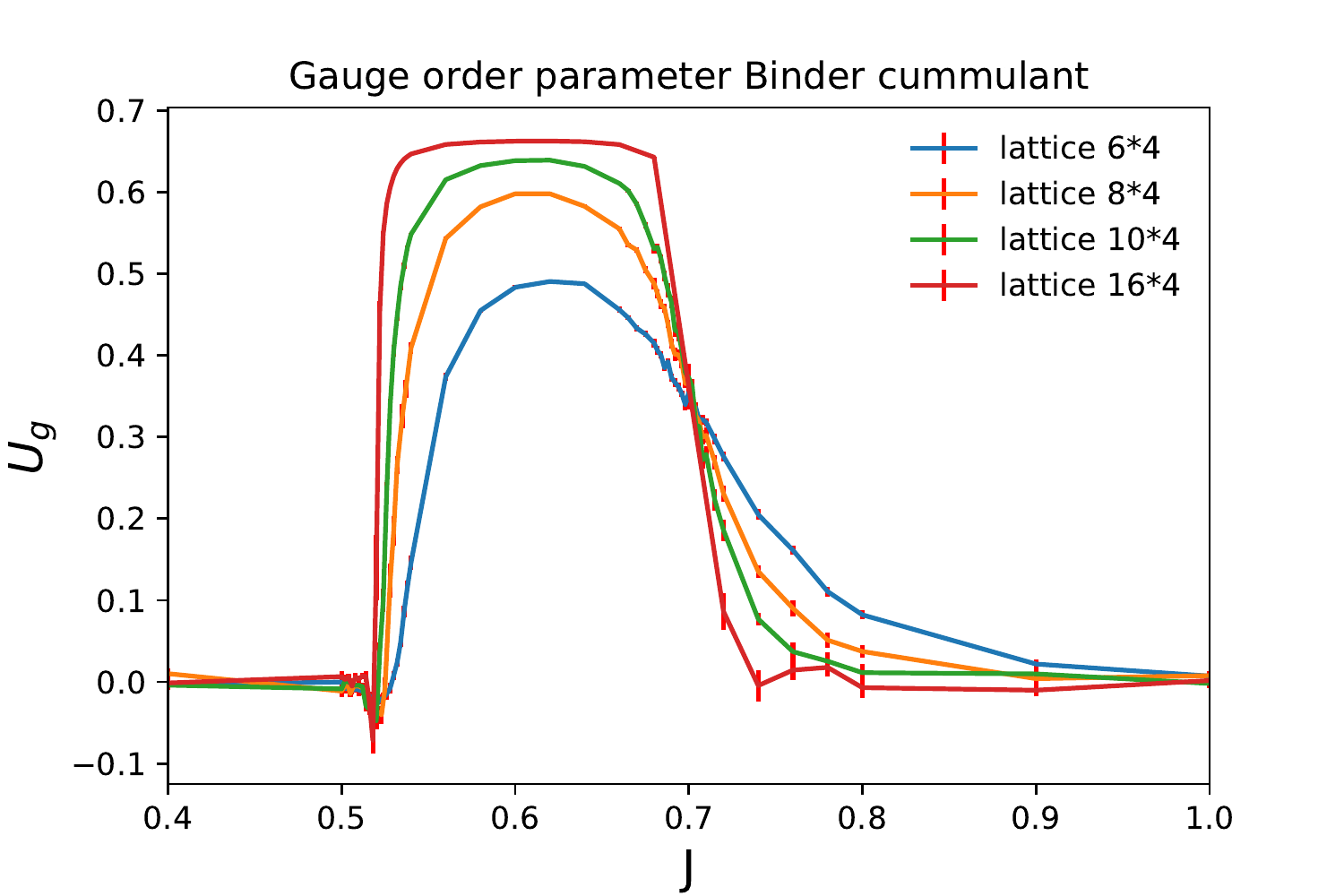} 
   \hspace*{1mm}\includegraphics[width=79mm]{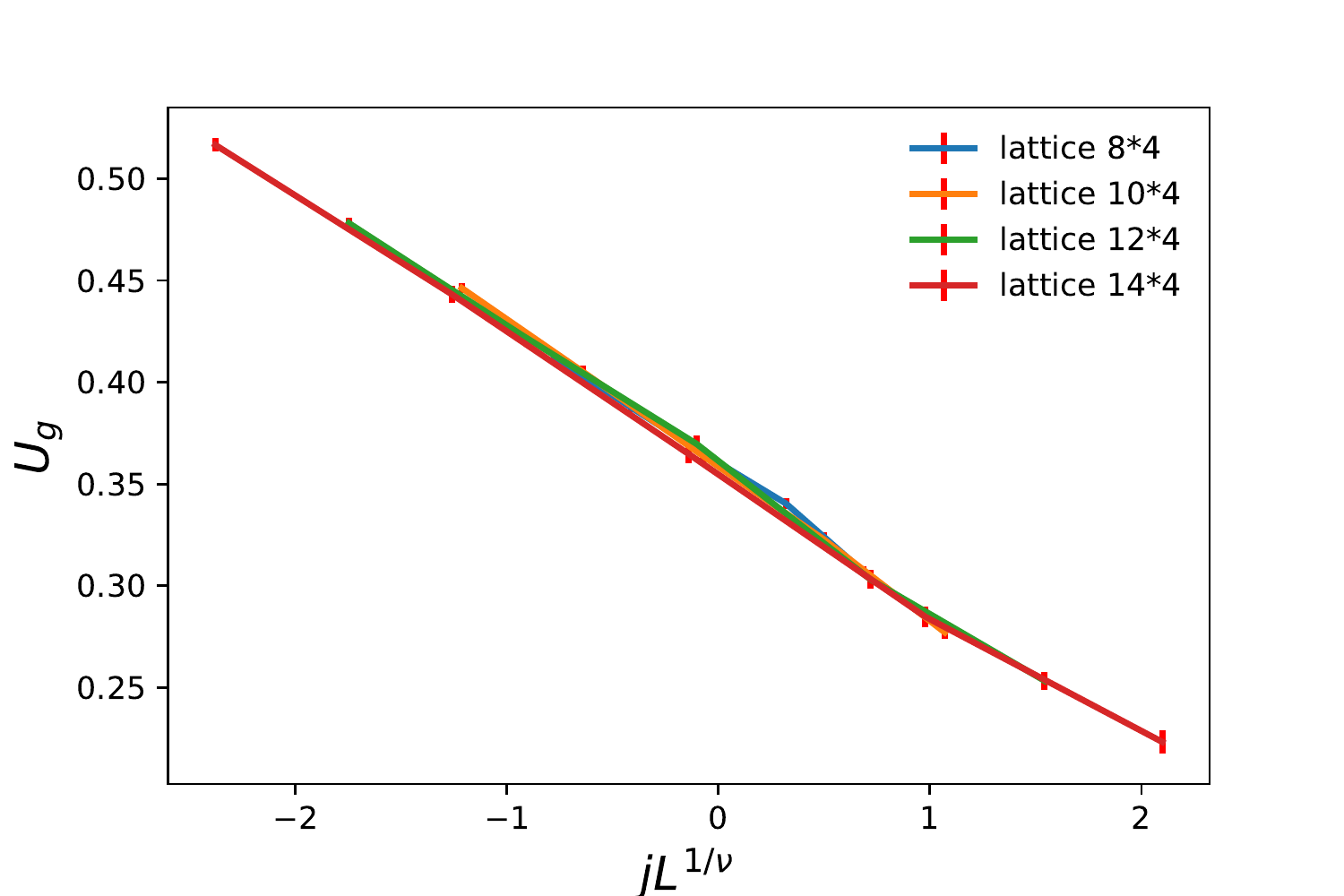}
   \hspace*{-8mm}
   \includegraphics[width=78mm]{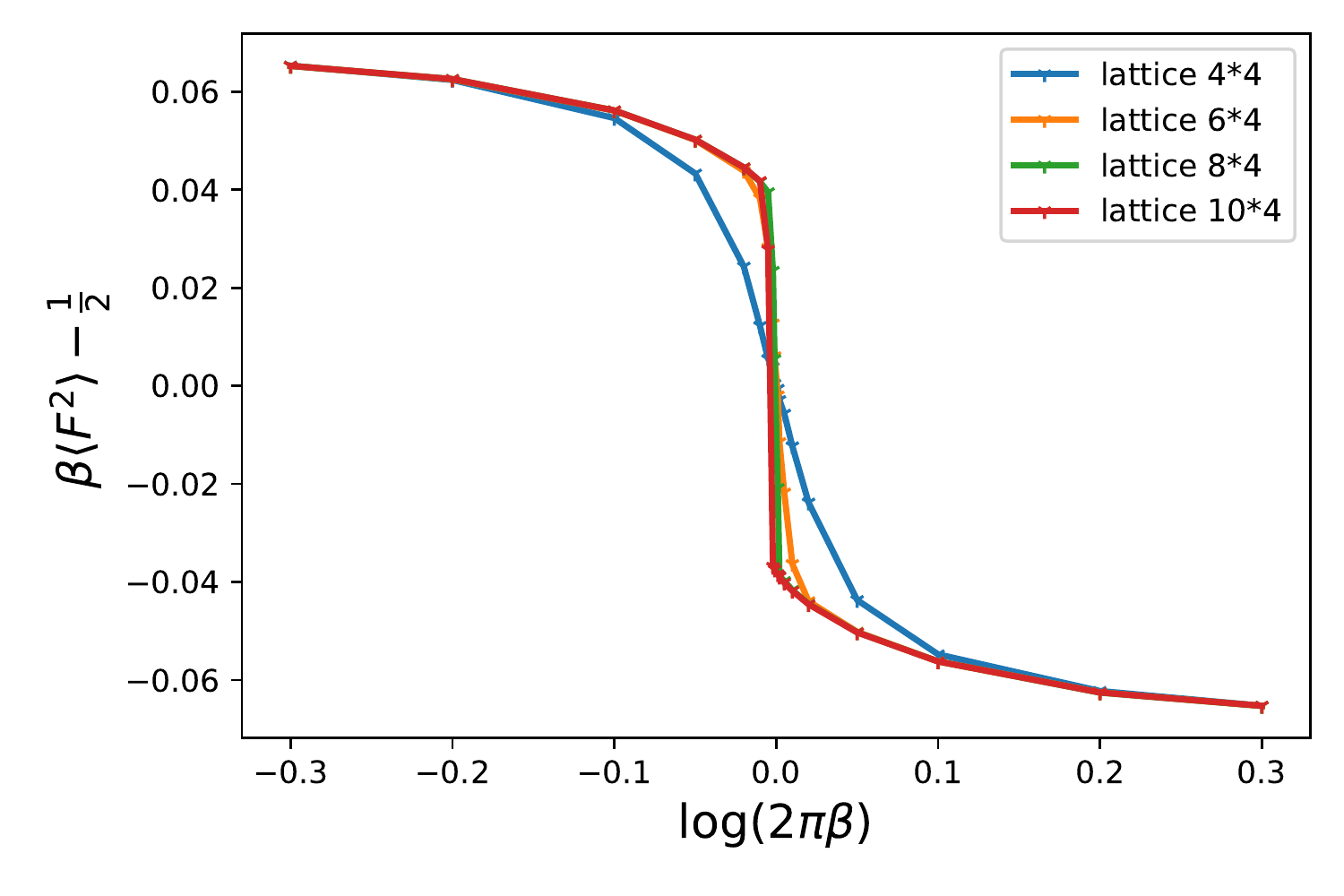} 
   \caption{Top row: Vacuum expectation value (left) and Binder cumulant (right) 
   for the gauge field order parameter $M_g$. We show the results at $\beta = \beta^* \equiv 1/2\pi$ as a function of 
   $J$ and compare the results for different volumes. Bottom left: The Binder cumulant 
   for the gauge field order parameter. For different volumes we 
   show the results at $\beta = \beta^*$ as a function of the rescaled coupling $j \, L^{1/\nu}$ with $j = J - J_2$ 
   and $\nu = 0.5$. Bottom right:  The gauge field order parameter as a function 
   of the rescaled gauge coupling $\ln (2\pi \beta)$, for values across the self-dual value $\beta^* \equiv 1/2\pi$.
   The matter field coupling here is set to $J = 0.6$.}
   \label{fig:overview}
\end{figure}

The top row of Fig.~\ref{fig:overview} shows the results for our gauge field observables at the critical 
coupling $\beta^* \equiv 1/2\pi$. The lhs.\ plot is the order parameter and the rhs.\ the corresponding 
Binder cumulant, both plotted as a function of the matter coupling $J$.
In the interval between $J \sim 0.52$ and $J \sim 0.7$ the order parameter develops 
a non-vanishing expectation value, while outside this interval it approaches 0 in the infinite volume 
limit. The corresponding Binder cumulant on the rhs.\ indicates that at 
$J_1 \sim 0.52$ we find a first order transition (the Binder cumulant develops a minimum at $J_1$), while the transition at 
$J_2 \sim 0.7$ turns out to be second order (Binder cumulants for different volumes intersect at the critical point). The matter field observables show essentially the same behaviour.

To locate the first order point near $J_1 \sim 0.52$ one may zoom into the region around $J_1$ and inspect 
the Binder cumulants for different volumes there. One observes, that for both the gauge and 
the matter cumulants a minimum forms, which for the largest three lattices and both observables is located at
the same point. We conclude that the system has a first order endpoint at  $J_1 = 0.518(2)$. 

For analyzing the second order transition at $J_2$ we use the finite size scaling formula 
$U \; \sim \; A \; + \; B  \, ( J - J_2) \, L^{1/\nu}$
for the Binder cumulant, where $A$ and $B$ are constants, $\nu$ is the critical exponent for the correlation length and we set 
$N_s = N_t = L$. The scaling formula implies that when plotted as a function of the rescaled coupling $j \, L^{1/\nu}$ with 
$ j = J - J_2$, the values of the Binder cumulant for different system sizes $L$ should collapse onto a single curve
(for correctly chosen values of $\nu$ and $J_2$). In the bottom left plot of 
Fig.~\ref{fig:overview} we show the result for the gauge field
Binder cumulant for a critical exponent $\nu = 0.5$. We zoom into the region near the second order transition point 
at $j = 0$ and for $J_2$ use the value where we find optimal collapse of the data, which is $J_2 = 0.700(1)$. 
Our first assessment of the transition at $J_2$ thus suggests that 
we have a second order transition at $J_2 = 0.700(1)$ with critical exponent $\nu = 0.5$, i.e., the mean field value. We are currently 
analyzing the scaling behavior of susceptibilities near $J_2$ in order to determine
the critical exponent $\gamma$. A first assessment indicates that also this exponent is compatible with the mean field value 
$\gamma = 1$.

We finally study our two order parameters as a function of the gauge coupling parameter $\beta$ in the vicinity 
of the self-dual point $\beta^* = 1/2\pi$. Values different from $\beta^*$ give rise to explicit breaking of self-duality
and thus setting $\beta \neq \beta^*$ amounts to introducing a symmetry breaking term, similar to the introduction of
an external magnetic field in a ferromagnet. To make this role more explicit we study the order parameters as a function of
$\ln (  2\pi \beta )$, which vanishes for $\beta = \beta^* = 1/2 \pi$ and is an odd function when interchanging $\beta$ and 
$\widetilde{\beta}$. We remark that for the gauge field order parameter we study the combination 
$\beta \langle F^2 \rangle_{\beta, J} - 1/2$ which vanishes at
the self-dual point $\beta = \beta^*$ and is an odd function of $\ln (  2\pi \beta )$. The bottom right plot in Fig.~\ref{fig:overview} nicely 
illustrates the behavior expected for an order parameter when the symmetry breaking coupling changes sign. 
We observe a finite jump when $\ln (  2\pi \beta )$ crosses 0, and curves corresponding to different 
lattice sizes quickly coincide with increasing volume. Thus we conclude that the line at $\beta = \beta^*$ between $J_1$ and $J_2$
is a first order line. 

\section{Discussion}

\noindent
In the project presented here we set out to study duality and self-duality for suitably discretized (Villain formulation) U(1) 
lattice gauge fields coupled to electric and magnetic matter. The system has a sufficiently rich duality structure, such that one may
study self-duality as a function of the matter coupling $J$. Of particular interest is the question whether self-duality may become
broken spontaneously and what the corresponding phase diagram looks like. 

Our analysis shows that for the self-dual gauge coupling $\beta^*$  
indeed self-duality becomes broken for an interval $[J_1,J_2]$ of matter couplings. Using suitable
order parameters and their Binder cumulants we conclude that we have a first order endpoint 
at $J_1 = 0.518(2)$ and a second order point at $J_2 = 0.700(1)$ with critical exponents 
that are compatible with the mean field values. Changing the gauge coupling $\beta$ away from the self-dual coupling 
$\beta^*$ is equivalent to introducing an explicit self-duality breaking term. Thus, when driving the gauge coupling 
through $\beta^*$ for some coupling $J \in [J_1,J_2]$ one expects a first order jump, which indeed is what we observe. 

The preliminary study presented here sheds light on the rich structure of duality and self-duality relations and 
non-perturbatively explores
their role for the phenomenology of quantum field theories. We remark that considerably more complex questions could 
be addressed, such as explicit breaking of self-duality by setting $J_e \neq J_m$, or adding additional self-dual terms that 
could alter the nature of the endpoints at $J_1$ and $J_2$. We are currently exploring some of these options.

\bibliographystyle{JHEP}

\bibliography{bibliography}

\end{document}